\documentstyle[12pt]{article}
\def\header{\begin{flushleft}
            ZU-TH 25/94\\August 1994
            \end{flushleft}}

\columnwidth\textwidth

\def\GSW_sign{}

\newcommand{\ba}{\begin{array}}
\newcommand{\ea}{\end{array}}
\newcommand{\bd}{\begin{displaymath}}
\newcommand{\ed}{\end{displaymath}}
\newcommand{\be}{\begin{equation}}
\newcommand{\ee}{\end{equation}}
\newcommand{\bea}{\begin{eqnarray}}
\newcommand{\eea}{\end{eqnarray}}

\def\bra{\langle}
\def\ket{\rangle}

\def\a{\alpha}
\def\b{\beta}
\def\g{\gamma}

\def\e{\epsilon}

\def\l{\lambda}
\def\m{\mu}
\def\n{\nu}
\def\G{\Gamma}

\def\L{\Lambda}
\def\s{\sigma}
\def\p{\pi}

\def\mv{m_{K^\ast}}

\def\to{\rightarrow}

\begin{document}
\thispagestyle{empty}
\header \vspace*{2cm} \centerline{\Large\bf Effects of new physics in the }
\centerline{\Large\bf  rare
  decays $B \to K \ell^+ \ell^-$ and $B \to K^\ast \ell^+ \ell^-$
 $\footnote{partially supported by
    Schweizerischer Nationalfonds.}$} \vspace*{2.0cm}
\centerline{\large C. Greub$^a$, A. Ioannissian$^b$, D. Wyler$^a$}
\vspace*{0.5cm} \centerline{\large $^a$ Universit\"at Z\"urich, 8057
  Z\"urich, Switzerland} \vspace*{0.5cm} \centerline{\large $^b$
 Yerevan Physics Institute, Yerevan, Armenia}
\vspace*{3cm} \centerline{\Large\bf Abstract} \vspace*{1cm} We
parametrize
phenomenologically  possible new physics effects
and
calculate their influence on the invariant dilepton mass spectrum and
the Dalitz plot for the decays
$B \to K \ell^+ \ell^-$ and $B \to K^\ast \ell^+ \ell^-$.
Especially the decay into $K^*$ yields a wealth of new information
on the form of the new interactions since the Dalitz plot is sensitive
to subtle interference effects. We also show how transversely polarized
$K^*$-mesons give additional information.

\newpage
\section{Introduction}
Rare decays of $B$-mesons, such as the recently observed
processes $B\rightarrow K^*\gamma$ and $ B \to X_s \g$
\cite{COR1,COR2} may become an
important tool for studying new forces beyond the standard
model. Indeed, many
authors have investigated the effects of multi-Higgs models,
supersymmetric theories, left-right symmetric models, etc. on
this process \cite{Hewett}. The
interest in these decays stems from the fact that they occur in
the standard model only through loops and are therefore particularly
sensitive to ``new physics''.

Whereas most of the past work has been concernend with the decay
$b\rightarrow s\gamma$, the leptonic transition $b\rightarrow s
\ell^+\ell^-$ is more sensitive to the actual form of the new
interactions since it not only allows to measure a total rate,
but also various kinematical distributions. More specifically,
while new forces will only change the normalization of $b\rightarrow
 s\gamma$ (or $B \to K^\ast \g$),
the interplay of the various operators will also change
the spectra in the decays with two leptons. Thus, different new physics
models which yield the same
 $b\rightarrow s\gamma$
 amplitude may
be distinguished by their characteristic invariant mass spectrum
of the lepton pair
in $b\rightarrow s \ell^+\ell^-$. Of course,
there is a spectrum for the  photon in the
inclusive process $B\rightarrow X_s \gamma$,
but it is essentially due to the wave function of the decaying
meson and therefore not of interest here.

In this note we will parametrize possible new physics
in a phenomenological way and evaluate their effects on the exclusive
decays  $B\rightarrow K \ell^+\ell^-$ and $B\rightarrow K^* \ell^+\ell^-$.
Similar results also follow for the inclusive lepton pair spectrum, but
since coherence effects tend to be more pronounced in exclusive decays
we have concentrated here on these. Of course, this introduces a
certain dependence on the hadronic model used. However, since
most models do not disagree
violently in the regions of interest and the crucial signs between the
form factors are quite safe, we feel that a reliable representation of the
effects is indeed possible. We have made
 use of a recent quark model
\cite {JA,JAW,JANEW} which gives rather satisfactory results. Needless
to say that it reproduces the heavy quark \cite{HQ} limit.


\section{Effective Hamiltonian}
The effects of both the new and the heavy standard model particles
can be summarized by an effective Hamiltonian; for
$b \to s \ell^+ \ell^-$ (and the corresponding exclusive decays) it
has the form \cite{GW}
\be
\label{heff}
{\cal H} = \frac{4 G_F \lambda_t}{\sqrt{2}} \, \sum_{i=1}^{10}
C_i(\m) \, O_i(\m) \quad ,
\ee
where $G_F$ denotes the Fermi constant and $\l_t = V_{tb} V_{ts}^\ast$
are products of CKM matrix elements. The
 $C_i(\mu)$ are coefficients calculated from processes
with heavy particles exchanged and $O_i$ are the following
operators
\bea
\label{basis}
O_1 &=& \left( \bar{s}_\a \g^\m L b_\a \right) \
        \left( \bar{c}_\b \g_\m L c_\b \right) \quad , \nonumber \\
O_2 &=& \left( \bar{s}_\a \g^\m L b_\b \right) \
        \left( \bar{c}_\b \g_\m L c_\a \right) \quad , \nonumber \\
O_7 &=& \frac{e}{16 \p^2} \, m_b \, (\bar{s}_\a \s^{\m \n} R b_\a)
         \, F_{\m \n} \quad , \nonumber \\
O_{7'} &=& \frac{e}{16 \p^2} \, m_b \, (\bar{s}_\a \s^{\m \n} L b_\a)
         \, F_{\m \n} \quad , \nonumber \\
O_8 &=& \frac{e^2}{16 \p^2} \, (\bar{s}_\a \g^\m L b_\a) \bar{e}
         \g_\m e \quad ,  \nonumber \\
O_{8'} &=& \frac{e^2}{16 \p^2} \, (\bar{s}_\a \g^\m R b_\a) \bar{e}
         \g_\m e \quad ,  \nonumber \\
O_9 &=& \frac{e^2}{16 \p^2} \, (\bar{s}_\a \g^\m L b_\a)
         \bar{e} \g_\m \g_5 e \quad , \nonumber \\
O_{9'} &=& \frac{e^2}{16 \p^2} \, (\bar{s}_\a \g^\m R b_\a)
         \bar{e} \g_\m \g_5 e \quad . \nonumber \\
\eea
The basis of ref. \cite{GW} has been supplemented by the primed
operators; the operators not written give small contributions.
In eq. (\ref{heff}),
the renormalization scale $\m$ is usually chosen to be
 $\m \simeq
m_b$ in order to avoid large logarithms in the matrix
elements of the operators $O_i$. Accordingly, the $C_i$ are shifted
from their perturbative value at $\m \simeq m_W$ by the
renormalization group equations.
As the analytic expressions for the coefficients $C_i(\m)$
in the standard model are given in many
places in the literature \cite{wilsoncoeff},
we only note here their numerical values
which correspond to $m_t=174$ GeV \cite{top}:
At the scale $\m=m_W$ one gets
\vspace{0.5cm}
\bea
\label{coeffmw}
C_1(m_W) &=& 0 \ , \ C_2(m_W) = -1 \ , \
C_7(m_W) = 0.196 \nonumber \\
C_{7'}(m_W) &=& \frac{m_s}{m_b} C_7(m_W)
\ , \ C_8(m_W) = -2.02 \nonumber \\
C_9(m_W) &=& 4.56 \ , \ C_{8'}(m_W)=C_{9'}(m_W)=0 \quad ,
\eea
while at $\m=m_b$ the non-vanishing coefficients are
\bea
\label{coeffmb}
C_1(m_b) &=& 0.226 \ , \ C_2(m_b) = -1.096 \ , \
C_7(m_b) = 0.310 \nonumber \\
C_{7'}(m_b) &=& \frac{m_s}{m_b} C_7(m_b)
\ , \ C_8(m_b) = -3.84 \ ,
\ C_9(m_b) = 4.56 \quad .
\eea
It is well-known that the operators $O_1$ and $O_2$ generate both
short  and long distance contributions
to $ b \to s \ell^+ \ell^-$ and to the corresponding exclusive decays.
The long distance contributions are mainly due to the $J/\psi$ and $\psi'$
resonances \cite{SAN,DES,Odonnel}.
These short and long distance contributions
can be easily built-in \cite{GW,AMM}
by replacing the coefficient
$C_8$ by $C_8^{eff}$, i.e.
\bea
\label{c8eff}
C_8^{eff}(m_b) &=& C_8(m_b) +
\left[
3 C_1(m_b) + C_2(m_b) \right] \times \nonumber \\
&& \times
\left[h(\hat{m}_c,\hat{s})
- \frac{3}{\a_{em}^2} \kappa \sum_{V_i=J/\psi,\psi'}
\frac{\p \G(V_i \to \ell \ell) M_{V_i}}{q^2-M_{V_i}+i M_{V_i}
\G_{V_i}} \right]
\quad ,
\eea
where $\hat{m}_c = m_c/m_b$ and $\hat{s} = q^2/m_b^2$ with
$q^2$ beeing the invariant mass squared of the lepton pair.
The short distance contributions are contained in the
function $h(\hat{m}_c,\hat{s})$ which reflects the one-loop
matrix element of the four-quark operators $O_1$ and $O_2$.
For $\hat{s} > 4 z^2$
\bea
\label{gfunim}
h(z,\hat{s}) &=& - \left\{
\frac{4}{9} \log z^2 - \frac{8}{27} - \frac{16}{9} \frac{z^2}{\hat{s}} +
\frac{2}{9} \sqrt{1-\frac{4z^2}{\hat{s}}} \, \left(
2 + \frac{4 z^2}{\hat{s}} \right) \right. \nonumber \\
&& \times \left.
\left[ \log \left|
\frac{1+\sqrt{1-4z^2/\hat{s}}}{1-\sqrt{1-4z^2/\hat{s}}} \right|
 + i \p
\right] \right\}
\eea
and for $\hat{s}<4z^2$
\bea
\label{gfunre}
h(z,\hat{s}) &=& - \left\{
\frac{4}{9} \log z^2 - \frac{8}{27} - \frac{16}{9} \frac{z^2}{\hat{s}} +
\right. \nonumber \\
&& \left.
+ \frac{4}{9} \sqrt{\frac{4z^2}{\hat{s}}-1} \, \left(
2 + \frac{4 z^2}{\hat{s}} \right)   \mbox{atan} \left(
\frac{1}{\sqrt{4z^2/\hat{s}-1}} \right) \right\} \quad .
\eea
The value of
$\kappa$ in eq. (\ref{c8eff})
 must be chosen such that  the combination
\[ \kappa \left[ 3 C_1(m_b) + C_2(m_b) \right] \approx -1 \]
in order to correctly reproduce the branching ratio \cite{AMM}
\[
\mbox{BR}(B \to J/\psi X \to \ell^+ \ell^- X) = \mbox{BR}(B \to J/\psi X)
\ \mbox{BR}(J/\psi \to \ell^+ \ell^-) \quad . \]

In this language, the physics beyond the standard model is
characterized by the
values of the Wilson coefficients
$C_i$ at the perturbative scale $m_W$. For instance,
supersymmetry can
lead to a $C_7$ which is opposite to its standard model value \cite{Susy},
whereas left-right symmetric models could give a range for $C_{7'}$
between minus and plus the standard model value of $C_7$
\cite{Hewett,Babu,AsIo}.
Models with leptoquarks may cause the other coefficients
to lie within
similar intervals. In order to limit ourselves to a reasonable set
of values in the examples, we chose to vary the coefficients between
minus and plus their standard model values. In addition, we require
that the rates for $B \rightarrow X_s \gamma$
and $B \to K^\ast \gamma$ do not change. For both reactions,
this implies that we must leave the
value of $(|C_7(m_b)|^2 + |C_{7'}(m_b)|^2)$
fixed within certain limits.

The running of the coefficients $C_{7'}$, $C_{8'}$ and $C_{9'}$
is simpler than the one of the unprimed coefficients since there
is no operator mixing with $O_2$.
In principle, new operators $O_{1'}$
and $O_{2'}$ with right-handed four fermion interactions should be
included; we assume them to be small, as indicated by
the high accuracy of the $V-A$ form in many weak decays, and
we neglect them. For the scaling we can simply write
in leading-log accuracy
\bea
\label{coeffprime}
C_{8'}(m_b) &=& C_{8'}(m_W) \nonumber \\
C_{9'}(m_b) &=& C_{9'}(m_W) \nonumber \\
C_{7'}(m_b) &=& \eta^{-16/23} C_{7'}(m_W) \quad ,
\eea
where $\eta=\a_s(m_b)/\a_s(m_W) \simeq 1.71$ for $m_b=5$ GeV.


\section{The decay $B \to K \ell^+ \ell^-$ }
The relevant hadronic matrix elements for the
decay $B \to K \ell^+ \ell^-$ associated with
the operators $O_i$ are parametrized as
in eqs. (2.11) and (2.12) of ref. \cite{JAW}, i.e.
\be
\label{eq211}
\bra p_K | \bar{s} \g_\m (1 \mp \g_5) b|p_B \ket =
F_+(q^2) P_\m + F_-(q^2) q_\m \quad ,
\ee
\be
\label{eq212}
\bra p_K | \bar{s} i \s_{\m \n} q^\n (1 \mp \g_5) b |p_B \ket =
\frac{1}{m_B+m_K} \left[ P_\m q^2 - (m_B^2 - m_K^2) q_\m \right] F_T(q^2)
\quad ,
\ee
where $P=p_B+p_K$ and $q=p_B-p_K$ and $\s_{\m \n}$ is defined as
$\s_{\m \n}=(i/2) [\g_\m ,\g_\n ]$.
The form factors $F_+$, $F_-$ and $F_T$ are written generically as
\be
\label{formgeneric}
F(q^2) = \frac{F(0)}{1-q^2/\L_1^2+ q^4/\L_2^4} \quad .
\ee
The parameters $\L_1$ and $\L_2$
characterizing the shape of each of
the form factors are given together
with $F(0)$ in table 1 which has been taken from ref. \cite{JAW}.
The full information on the decay depends on two kinematical variables
which we choose to be $\hat{s}=q^2/m_B^2$ and $x=E_\ell/m_B$, where
$q^2$ is
the invariant mass of the lepton pair and $E_{\ell}$ is the
energy of the negatively charged lepton measured in the rest frame
of the $B$ meson. Using the notation of ref. \cite{IWise}
we get the double differential decay width
\be
\label{bk}
\frac{d^2\G(B \to K \ell^+ \ell^-)}{dx d\hat{s}} =
\frac{G_F^2 m_B^5 |\l_t|^2}{8 \p^3}  \,  \frac{\a_{em}^2}{16 \p^2} \, \b \,
 [4x(x_m-x)-\hat{s}(1-2x)]
\quad ,
\ee
where
\be
\label{betabk}
\b = \left| (C_8^{eff}+C_{8'}) F_+ -
\frac{2 m_b(C_7+C_{7'})}{m_B+m_K} F_T \right|^2 +
\left| (C_9+C_{9'}) F_+ \right|^2 \quad .
\ee
The variables $x$ and $\hat{s}$ vary in the range
\be
\label{xsrange}
\frac{4 m_\ell^2}{m_B^2} \le \hat{s} \le \frac{(m_B - m_K)^2}{m_B^2}
\quad , \quad
\frac{\hat{s}+ 2 x_m - \sqrt{\phi}}{4} \le x \le
\frac{\hat{s}+ 2 x_m + \sqrt{\phi}}{4} \quad ,
\ee
with
\be
\label{xmphidef}
x_m = \frac{m_B^2-m_K^2}{2 m_B^2} \quad , \quad
\phi = (\hat{s} + 2 x_m)^2 - 4 \hat{s}.
\ee
Integrating eq. (\ref{bk}) over
the variable $x$ in the range specified in eq. (\ref{xsrange}) yields
the invariant mass distribution of the lepton pair
\be
\label{bk1}
\frac{d \G(B \to K \ell^+ \ell^-)}{d\hat{s}} =
\frac{G_F^2 m_B^5 |\l_t|^2}{96 \p^3}  \,  \frac{\a_{em}^2}{16 \p^2} \, \b \,
 \phi^{3/2}
 \quad .
\ee

We notice the relative sign in eq. (\ref{betabk})
between the terms involving the $C_7$ and $C_8$ coefficients
which leads to a potentially
interesting interference effect.
Since $F_+$ and $F_T$ are calculated in a specific model, one might
worry that the sign is strongly model dependent. However, in the
approximation where the quarks are on-shell, the Gordon
decomposition  confirms the sign.

In Fig. 1 we show the dilepton invariant mass
distribution for
the standard model and for a model where $C_8$
has the opposite sign from
the standard model value. The difference between
the two curves is
especially noticeable near the resonances,
a fact which is easily
explained by eq. (\ref{c8eff}) (see also eq. (\ref{betabk})).
On the other hand,
changing the sign of
$C_7$ does not alter the standard model picture
very much, since the
contributions from the $C_7$ terms in eq. (\ref{betabk}) are small.
{}From the form of $\beta$ we see that other variations of the
coefficients will not give dramatic changes of the spectra and Dalitz
plot.

In the numerical evaluations we take
$m_b=m_B=5.28$ GeV in all formulas
except in the Wilson coefficients, where we used $m_b=5$ GeV.
Furthermore,
in all the plots we divided the spectra by the total $B$-meson decay width,
estimated to be \cite{Paschos}
\be
\label{gammatot}
\G_{tot} = \frac{f \, m_b^5 G_F^2 |V_{cb}|^2}{192 \p^3} \quad ,
\ee
with $f \approx 3$.


\section{The decay $B \to K^\ast \ell^+ \ell^-$ }
Next, we consider the more interesting decay
  $B \to K^\ast \ell^+ \ell^-$.
Again, the relevant hadronic matrix elements are parametrized as
in eqs. (2.18) and (2.19) of ref. \cite{JAW}, i.e.
\bea
\label{eq218}
 && \bra p_{K^\ast} | \bar{s}
\g_\m (1 \mp \g_5) b|p_B \ket =
\frac{1}{m_B+m_{K^\ast}} \,
\left[-i V(q^2)
 \epsilon_{\m \n \a \b} \e^{\ast \n} P^\a
q^\b \right. \nonumber \\
&& \left.
\pm A_0(q^2) (m_B^2-m_{K^\ast}^2) \e^\ast_\m \pm A_+(q^2)
 (\e^\ast P) P_\m \pm
A_-(q^2) (\e^\ast P) q_\m \right]
\eea
\bea
\label{eq219}
&& \bra p_{K^\ast} | \bar{s}
i \s_{\m \n} q^\n (1 \pm \g_5) b |p_B \ket =
-i g(q^2)
\epsilon_{\m \n \a \b} \e^{\ast \n} P^\a q^\b \pm \nonumber \\
&& a_0(q^2) \,
(m_B^2 - m_K^2) \left[ \e^\ast_\m - \frac{1}{q^2} (\e^\ast q) q_\m
\right] \pm
a_+(q^2) \,
(\e^\ast P) \left[ P_\m - \frac{1}{q^2} (Pq) q_\m \right]
\quad ,
\eea
where
$P$ and $q$ are now defined as $P=p_B + p_{K^\ast}$ and
$q=p_B - p_{K^\ast}$.
The $\epsilon$-tensor is taken in the Bjorken-Drell convention
(i.e. $\epsilon_{0123}=1$) contrary to ref. \cite{JAW}, which explains the
additional minus sign of the terms proportional to the $\epsilon$-tensor.
The various form factors in eqs. (\ref{eq218}) and  (\ref{eq219})
are written as in eq.
(\ref{formgeneric}) and the parameters are  given in
table 1.
 With the notation of ref. \cite{IWise}
we get
\bea
\label{bkstar}
\frac{d^2\G(B \to K^\ast \ell^+ \ell^-)}{dx d\hat{s}} &=&
\frac{G_F^2 m_B^5 |\l_t|^2}{4 \p^3} \, \frac{\a_{em}^2}{16 \p^2} \,
\left[
\frac{\hat{s}}{m_B^2} \, \a + 2 \b [4x(x_m-x)- \hat{s}(1-2x)]
 \right. \nonumber \\
&& \left. \hspace{1cm} - 2 \g \hat{s} (
x_m -2x + \hat{s}/2)
\right] \quad ,
\eea
where
\bea
\label{parambkstar}
\a &=& |G_A^0|^2 + |F_A^0|^2 + 4 m_B^2 p_s^2
\left(|G_V|^2 + |F_V|^2 \right) \nonumber \\
\b &=& \frac{|G_A^0|^2 + |F_A^0|^2}{4 \mv^2} - m_B^2 \hat{s}
 \left(|G_V|^2 + |F_V|^2 \right)
+ \frac{m_B^2}{\mv^2} p_s^2
\left(|G_A^+|^2 + |F_A^+|^2 \right)
+ \nonumber \\
&& \frac{1}{2} \left[ \frac{m_B^2}{\mv^2}(1-\hat{s})-1 \right] \mbox{Re}
(G_A^0 G_A^+ + F_A^0 F_A^+) \nonumber \\
\g &=& -2 \mbox{Re} (G_A^0 F_V + F_A^0 G_V ) \quad ,
\eea
and
\[
p_s^2 = \phi \frac{m_B^2}{4} \quad .  \]
The functions $G_V$, $F_V$, $G_A^0$, $F_A^0$, $G_A^+$ and $F_A^+$ are
combinations of form factors and Wilson coefficents; they read
\bea
\label{gi}
G_V &=& \frac{(C_8^{eff}+C_{8'}) V}{2(m_B+\mv)} - (C_7+C_{7'})
\frac{m_b}{q^2} g  \nonumber \\
F_V &=&  \frac{(C_9+C_{9'})V}{2(m_B+\mv)} \nonumber \\
G_A^0 &=& \frac{(C_8^{eff}-C_{8'})(m_B-\mv) A_0}{2} - (C_7-C_{7'})
\frac{m_b}{q^2}(m_B^2-\mv^2) a_0 \nonumber \\
F_A^0 &=& \frac{(C_9 - C_{9'})}{2} (m_B - \mv) A_0 \nonumber \\
G_A^+ &=& \frac{(C_8^{eff}-C_{8'}) A_+}{2(m_B+\mv)} -(C_7 - C_{7'})
\frac{m_b}{q^2} a_+ \nonumber \\
F_A^+ &=&  \frac{(C_9 - C_{9'})A_+}{2(m_B+\mv)} \quad .
\eea
The range of the variables $x$ and $\hat{s}$
and the definition of $x_m$ are obtained from eqs.
(\ref{xsrange}) and (\ref{xmphidef})
by the obvious replacement $m_K \to \mv$.

Integration over the variable $x$ leads to the invariant mass distribution
of the lepton pair
\be
\label{bkstar1}
\frac{d \G(B \to K^\ast \ell^+ \ell^-)}{d\hat{s}} =
\frac{G_F^2 m_B^5 |\l_t|^2}{8 \p^3} \, \frac{\a_{em}^2}{16 \p^2} \,
\sqrt{\phi} \,
\left[
\frac{\hat{s}}{m_B^2} \, \a + \frac{\b}{3} \, \phi \right] \quad .
\ee

Taking into account only transversely polarized $K^\ast$ mesons in the
final state, the analogous
expression for the double differential decay width is
\bea
\label{bkstart}
\frac{d^2\G^T(B \to K^\ast \ell^+ \ell^-)}{dx d\hat{s}}  &=&
\frac{G_F^2 m_B^5 |\l_t|^2}{4 \p^3} \, \frac{\a_{em}^2}{16 \p^2} \,
\left[
\frac{\hat{s}}{m_B^2} \, \a^T + 2 \b^T [4x(x_m-x)-\hat{s}(1-2x)]
\right. \nonumber \\
&& \hspace{-2.0cm} \left. - 2 \g^T \hat{s} (
x_m -2x + \hat{s}/2) - \frac{\delta^T}{2 m_B^2} \hat{s} \left(
\frac{4x (x_m-x)- \hat{s}(1-2x)}{(x_m+\hat{s}/2)^2-\hat{s}}
\right) \right]
 \quad ,
\eea
where
\be
\label{parambkstart}
\a^T = \a \ , \ \b^T = - m_B^2 \hat{s} (|G_V|^2 + |F_V|^2) \ , \
\g^T = \g \ , \
\delta^T = |G_A^0|^2 + |F_A^0|^2 \ .
\ee
The corresponding invariant mass distribution of the lepton pair
reads
\bea
\label{bkstart1}
\frac{d \G^T(B \to K^\ast \ell^+ \ell^-)}{d\hat{s}} &=&
\frac{G_F^2 m_B^5 |\l_t|^2}{8 \p^3} \, \frac{\a_{em}^2}{16 \p^2} \,
\sqrt{\phi} \, \left[
 \frac{\hat{s}}{m_B^2} \, \a^T + \frac{\b^T}{3} \, \phi -
\right. \nonumber \\
&& \left.
\frac{\delta^T}{12 m_B^2} \hat{s}
\frac{\phi}{(x_m+\hat{s}/2)^2-\hat{s}}
\right]
\quad .
\eea

In Figs. 2 to 4 we show the invariant mass distribution
and
the Dalitz
plots for  three sets of values of the couplings,
all of which give the
same rate for the decay $B \rightarrow X_s \gamma$.
In Fig. 2  we illustrate the standard model prediction (Fig. 2a) and
the case where the sign of $C_8$ is reversed (Fig. 2b); this
again leads to the
marked deviation around the resonances. In addition,
we see from eqs. (\ref{parambkstar}) and (\ref{bkstar1})
that
reversing the sign of $C_9$ does not change the
invariant
mass distribution; however, this change
influences the Dalitz plot significantly as illustrated in Fig 3:
While the standard model distribution in Fig. 3a
is populated more for large $x$-values, Fig. 3c shows
more points for small values of $x$.

Finally, from Fig. 4 we see that the transverely polarized
$K^*$ mesons accentuate this asymmetry in the
Dalitz plot and can therefore
be used to fix the sign of the coefficient $C_9$.


\section{Conclusions}
In this paper we have investigated the effects of new physics on the
rare decays $B \to K \ell^+ \ell^-$ and $B \to K^\ast \ell^+ \ell^-$.
Whereas the new physics influence only the rate
in the decay $B \rightarrow X_s \gamma$,
they change significantly the
various spectra in the leptonic decays. Thus,
while certain models
give the same rate for  $B \rightarrow X_s \gamma$, they
can be distinguished by their dilepton
spectra. In addition, the Dalitz plot carries further
information (compare
Figs. 3 and 4). Furthermore, looking only at the
transversely polarized $K^*$ yields improved information; as an
example we have investigated the coefficient $C_9$.

Of course, the applicability of our results will depend on the number
of avaliable decays. It is expected that at the hadronic $B$ facilities
at LHC the required number (about 100
for the dilepton spectrum and
1000 for the Dalitz plot) can be analyzed. Then, these
decays will be a truly powerful tool towards understanding new physics.

\vspace{1cm}
Noted added:
A. Ali, G. Giudice and T. Mannel have investigated in a similar spirit the
inclusive dilepton decays (private communication and \cite{AGM}).

\section{Acknowledgements}
We thank P. Schlein and P. Sphicas for informative discussions.
One of the authors (A. I.) is grateful to the staff of the Institute of
Theoretical Physics of Zurich University for the warm hospitality.


\vspace*{\fill}
\subsection*{Figure captions}
\begin{description}

\item [Fig.\,1:] Invariant dilepton mass distribution for
the decay $B \to K \ell^+ \ell^-$. In Fig. 1a the prediction within
the standard model is given; Fig. 1b shows the result for a
model where $C_8(m_W) = 2.02$
(opposite value of the standard model;
 see eq. (\ref{coeffmw})) and the other
coefficients unchanged.

\item [Fig.\,2:] Same as Fig. 1, but for the decay
$B \to K^\ast \ell^+ \ell^-$. The solid (dashed) line takes into
account all (only the transverse) polarizations of the $K^\ast$-meson.

\item [Fig.\,3:] Dalitz plot
(compare eq. (\ref{bkstar}))
for the decay $B \to K^\ast \ell^+ \ell^-$
taking into account all polarizations of the $K^\ast$-meson. We have
used 8000 points in the scatter plots.
The distribution in Fig. 3a is for the standard model values of the
coefficients while Fig. 3b has the sign of $C_8(m_W)$ reversed.
In Fig. 3c we used  the standard model values of the the
coefficients, except that the sign of $C_9(m_W)$ has been reversed.

\item [Fig.\,4:] Dalitz plot for transversely polarized
$K^\ast$-mesons. Fig. 4a is for the standard model values of the
coefficients and Fig. 4b has the sign of $C_9(m_W)$ reversed.

\end{description}


\section*{Tables}
\begin{table}[h]
\begin{center}
\begin{tabular}{|r|r|r|r|r|r|r|r|r|}
\hline
          & $F_+$ & $F_T$ & $V$ & $A_+$ & $A_0$ & $g$ & $a_+$ & $a_0$\\
\hline
 $F(0)$           &  0.30 & $-0.30$ & $-0.35$ & 0.24 & $-0.37$ & 0.31 & $-0.31$
& 0.31
\\ \hline
 $\L_1$ (GeV)     &  4.06& 4.08 & 4.05  &  4.39& 5.92 & 4.05 & 4.37 & 6.40
\\ \hline
 $\L_2$ (GeV)     &  5.51 & 5.50& 5.45&5.81&8.54&5.44&5.84& 8.84
\\ \hline
\end{tabular}
\caption{Values of the various $F(0)$, $\L_1$ and $\L_2$}
\end{center}
\end{table}

\enddocument